\documentclass[12pt]{article} 

\newcommand{\re}{\mathrm{e}} 
\newcommand{\ri}{\mathrm{i}}

\RequirePackage{amssymb}
                      
\begin{document}
\begin{center}
{\bf Coherent distributions for the rigid rotator} \\[2cm] 
M. Grigorescu \\[3cm]  
\end{center}
\noindent
$\underline{~~~~~~~~~~~~~~~~~~~~~~~~~~~~~~~~~~~~~~~~~~~~~~~~~~~~~~~~
~~~~~~~~~~~~~~~~~~~~~~~~~~~~~~~~~~~~~~~~~~~}$ \\[.3cm]
Coherent solutions of the classical Liouville equation for the rigid rotator 
are presented as positive phase-space distributions localized on the Lagrangian 
submanifolds of Hamilton-Jacobi theory. These solutions become Wigner-type 
quasiprobability distributions by a formal discretization of the left-invariant 
vector fields from their Fourier transform in angular momentum. The results are  
consistent with the usual quantization of the anisotropic rotator, but the expected 
value of the Hamiltonian contains a finite "zero point" energy term. It is shown 
that during the time when a quasiprobability distribution evolves according to  
the Liouville equation,  the related quantum wave function should satisfy the 
time-dependent Schr\"odinger equation.  
\\
$\underline{~~~~~~~~~~~~~~~~~~~~~~~~~~~~~~~~~~~~~~~~~~~~~~~~~~~~~~~~~~~~~
~~~~~~~~~~~~~~~~~~~~~~~~~~~~~~~~~~~~~~~}$ \\
%e-print arXiv:1504.04832 \\[3cm]
\newpage
\section{Introduction}  
The models of rigid rotation concern finite many-particle systems having a well-defined 
instantaneous intrinsic frame. The space $Q$ of static configurations is in this case the 
group $SO(3, {\mathbb R})$ while dynamics near ground-state energy can be described as a Hamiltonian 
flow on the phase-space $M=T^*SO(3, {\mathbb R})$ \cite{jem}. \\ \indent
For the atomic nuclei such an intrinsic frame is defined by a self-consistent deformed mean-field.
The residual interactions between protons and neutrons can be taken into account by 
models of two coupled rotators\footnote{The term "rotator" was preferred here for being 
closer to "rotating body" than the common form "rotor". Though, there are references in which
"rotator" designates only a linear rotating object ($I_1=I_2 >0$, $I_3=0$).}, applied to study 
the low-energy isovector magnetic  excitations \cite{m1,trm}. Similar considerations for the isospin 
degree of freedom have  been presented in \cite{cpn1, cpn2}.  \\ \indent
Distributions on  $T^*SO(3, {\mathbb R})$ appear in the treatment of statistical 
ensembles of microscopic rotators. Such ensembles are particularly interesting because unlike
the space $Q={\mathbb R}^3$ of the translation coordinates, the spaces         
$SO(3, {\mathbb R})$ of the intrinsic rotations are physically distinct and of finite 
volume ($v_R=8 \pi^2$). For the polyatomic molecules the moments of inertia are large, and 
although the statistical weight may include the nuclear spin degeneracy \cite{landau}, the 
thermal equilibrium is described by the classical Boltzmann distribution. At low 
temperatures the partition function is calculated using the spectrum of the quantum 
Hamiltonian, which is well-known up to a controversial "zero-point" energy term. Proposed 
during the early days of the quantum theory \cite{ei} to explain the specific heats of 
diatomic gases, finite ground state-energy terms have been retrieved also in the 
more recent years, for the rigid rotator from geometrical considerations \cite{gc1,gqa,gc2},
and for the simple (linear) rotator by using reduced Wigner quasiprobability distributions \cite{wf1}. 
Moreover, by adapting the known Wigner transform \cite{wf} from ${\mathbb R}^3$ to $SO(3, 
{\mathbb R})$, an additional "quantum potential" term was obtained \cite{wf2}.  \\ \indent
In this work Wigner-type distributions for the rigid rotator will be introduced along the 
lines of \cite{cpw, cdq2}, by discretizing the Fourier transform in momentum of the 
"action waves" on  $T^*SO(3, {\mathbb R})$. The classical dynamics of the rigid rotator is 
presented in Section 2, both as a Hamiltonian system on $T^*SO(3, {\mathbb R})$ and as 
geodesic motion on $SO(3, {\mathbb R})$. The coherent solutions of the classical Liouville 
equation, provided by the "action waves" of the Hamilton-Jacobi theory, are presented in 
Section 3. In Section 4 it is shown that a formal discretization of the angular derivatives 
in the "action waves" provides Wigner-type  quasiprobability distributions on $T^*SO(3, {\mathbb R})$ 
which are consistent with the usual quantum treatment. However, for the expected value of the 
Hamiltonian, a finite "zero point" energy is obtained. Conclusions are summarized in Section 5.  
     
\section{The classical  equations of motion} 
The group $SO(3, {\mathbb R})$ consists of the $3 \times 3$ real, orthogonal, 
unimodular matrices, 
\begin{equation}
SO(3, {\mathbb R}) =  \{ {\cal R}~;~  {\cal R}^T {\cal R} = I, 
{\rm det } {\cal R} = 1   \} ~~. \label{gso3}
\end{equation}
This group is a compact manifold, such that every rotation ${\cal R}$ can be
specified by the versor  ${\bf g}$  of the rotation axis and the angle $\gamma$ of 
rotation around this axis, which means by the vector $\gamma {\bf g} \in 
{\mathbb R}^3~,~\vert {\bf g} \vert =1, ~\gamma \in [0, \pi]$. The parametrization of 
${\cal R}$ in terms of $\gamma {\bf g}$ and of the Euler angles $\varphi, \theta, \psi$ 
 is presented in Appendix 1. 
 \\ \indent
To describe the rigid rotator, the ``laboratory'' frame versors ${\bf e}_1, {\bf e}_2, 
{\bf e}_3$ are supposed to be fixed, and a rotated frame ${\bf e}_k' = 
\sum_i {\cal R}^T_{ki} {\bf e}_i$ is intrinsically attached to each orientation 
in ${\mathbb R}^3$ of the rotator, such that its configuration space  can be 
identified with $SO(3, {\mathbb R})$.  
\\ \indent
An integrable basis on  $TSO(3, {\mathbb R})$ is provided by $\partial_\varphi, \partial_\theta,
\partial_\psi$, but the intrinsic description corresponds to the local 
 basis $Z_1,Z_2,Z_3$, of left-invariant vector fields (Appendix 1). This is related to 
a local basis $\zeta_k$ of one-forms on $T^* SO(3, {\mathbb R})$,
$$ 
 \zeta_1 = \sin \theta \sin \psi d \varphi + \cos \psi d \theta  ~~,~~ 
$$
\begin{equation}
 \zeta_2 = \sin \theta \cos \psi d \varphi - \sin \psi d \theta   ~~,~~
\end{equation}
$$
\zeta_3 = \cos \theta d \varphi + d \psi~~, 
$$
such that $\zeta_i (Z_j) \equiv \langle \zeta_i,  Z_j \rangle = \delta_{ij}$. 
It is easy to check that 
\begin{equation}
d \zeta_1 = -  \zeta_2 \wedge \zeta_3~,~ d \zeta_2 = -  \zeta_3 \wedge
\zeta_1~,~ d \zeta_3 = -  \zeta_1 \wedge \zeta_2 ~~,
\end{equation}
and $\zeta_1 \wedge \zeta_2 \wedge \zeta_3 = - \sin \theta d \varphi \wedge
d \theta \wedge d \psi $. Thus, if the tangent to the trajectory ${\cal R}(t)$ 
is expressed in the form $X_\tau = \dot{\varphi} \partial_\varphi + \dot{\theta} 
\partial_\theta + \dot{\psi} \partial_\psi$, then $\zeta_i(X_\tau) = \omega'_i$ are the 
intrinsic components of the angular velocity. \\ \indent
The 1-forms $\zeta$ can be used to define a symmetric 2-form on $TSO(3, {\mathbb R})$, 
$B = \sum_k I_k \zeta_k \otimes \zeta_k$,
\begin{equation}
B(X,Y) = \sum_k I_k \zeta_k(X) \zeta_k(Y)~~,~~ X,Y \in TSO(3, {\mathbb R})~~.
\end{equation}
 If $I_k>0$ are the intrinsic moments of inertia (constants), then the kinetic 
energy of the rigid rotator can be written in the form $T = B(X_\tau, X_\tau)/2$. By the
Legendre transform ${\mathbb L}$ \cite{jem}, the 2-form $B$ provides a map
${\mathbb L} :TSO(3, {\mathbb R}) \rightarrow T^*SO(3, {\mathbb R}) $,  
$X \rightarrow {\mathbb L}_X$, ${\mathbb L}_X (Y) \equiv B(X,Y)$. 
In particular ${\mathbb L}_{X_\tau} = \sum_k \rho_k \zeta_k$, with $\rho_k =I_k \omega'_k$,
such that the kinetic energy $T$ also defines the classical Hamilton function 
$H = \sum_k \rho_k^2/2I_k$. 
  \\ \indent
Let $\Theta = \sum_k \rho_k \zeta_k$ be the canonical 1-form on 
$M=T^* SO(3, {\mathbb R})$, and $\Omega = - d \Theta $ the symplectic form. The 
dynamics of the rigid rotator can be described either as a Hamiltonian flow on 
$(M, \Omega)$, induced by $H$, or as geodesic motion on $SO(3, {\mathbb R})$ for 
the metric $B$. 
\subsection{The Hamiltonian approach} 
In classical mechanics, the  observables are represented within the set 
${\cal F} (M)$ of the smooth real functions on the symplectic manifold $(M, \Omega)$. 
Let $X_f$ be the vector field provided by  $i_{X_f} \Omega = d f$, $f \in {\cal F} (M)$, 
where $i_{X_f} \Omega$ denotes the inner product between  $X_f$ and $\Omega$.   
The set ${\cal F}(M)$  becomes a Lie algebra  with respect to the Poisson bracket
$\{ * , * \}$, 
\begin{equation}
\{ f,g \} = \langle df , X_g \rangle =  \omega(X_f,X_g) = - {\rm L}_{X_f} g,~~f,g 
\in {\cal F}(M)~~,
\end{equation}
in which  ${\rm L}_X$ denotes the Lie derivative with respect to  $X$. In the case 
$M= T^*SO(3, {\mathbb R})$, $\Omega = - d \Theta$, we get
\begin{equation}
\{ f,g \} = \sum_k (Z_k f \partial_{\rho_k} g -  Z_k g \partial_{\rho_k} f) - 
\vec{\rho} \cdot ( \vec{\partial}_\rho f \times \vec{\partial}_\rho g)~~, \label{pb} 
\end{equation}
where $\partial_{\rho_k} \equiv \partial / \partial \rho_k$, $\vec{\rho} \equiv ( 
\rho_1, \rho_2, \rho_3)$ and $f,g \in {\cal F}(M)$ are functions $f({\cal R},\vec{\rho})$, 
of ${\cal R}  \in SO(3, {\mathbb R})$ and  $\vec{\rho} \in {\mathbb R}^3$.
In particular, for the angular momentum components  $\rho_i$, in the intrinsic frame,  
and, respectively,  $l_i =\sum_k {\cal R}_{ik} \rho_k$, in the laboratory frame, we get 
\begin{equation}
\{ \rho_i, \rho_j \} = -  \epsilon_{ijk} \rho_k   ~~,~~  \{ l_i, l_j \} = \epsilon_{ijk} l_k~~. \label{pbm} 
\end{equation}
Thus, $\rho_i$ and $l_j$  are not usual canonical  momenta. A local set of canonical coordinates 
on $T^*SO(3, {\mathbb R})$  is presented in Appendix 2.    
\\ \indent
The change of sign in (\ref{pbm}) is due to the angular derivatives in (\ref{pb}), and for a 
spherical rotator having a magnetic moment $\gamma_m {\bf l}$, ($\gamma_m$ is the gyromagnetic ratio), 
placed in an external magnetic field  ${\bf B}$, without damping, from $\dot{\bf l} = 
\{ {\bf l} , H_0 \}$, $H_0= {\bf l}^2 /2I  - \gamma_m {\bf l}  \cdot {\bf B}$,  we get the 
Bloch equations, $ \dot{\bf l} = \gamma_m {\bf l} \times {\bf B}$. 
\\ \indent
In the case of a deformed rotator described by $H = \sum_k \rho_k^2/2I_k$
we may consider intrinsic Hamiltonian vector fields of the form 
$X_H = \sum_k \omega'_k Z_k + \dot{\rho}_k  \partial_{\rho_k}$, and 
$i_{X_H} \Omega = d H$ reduces to the Euler equations 
\begin{equation}
\omega_k'= \rho_k /I_k~~,~~ \dot{\vec{\rho}} = \vec{\rho} \times \vec{\omega}' ~~.  \label{eul} 
\end{equation} 
Because $\{ \rho_i, l_j \} =0$, we get also $\{ H, {\bf l} \} =0$, such that  
$l_i =\sum_k {\cal R}_{ik} \rho_k$ are constants and $\vec{\rho}$ from (\ref{eul}) evolves on the 
coadjoint orbit of  ${\bf l}$.  

\subsection{The geodesic approach}   
In the geodesic approach we take $B$ as a left-invariant metric on $TSO(3, {\mathbb R})$,
${\rm L}_Y B =0$. A curve ${\cal R}(t)$ on $SO(3, {\mathbb R})$ is a geodesic with 
respect to $B$ if its tangent field $X_\tau$ is "autoparallel", in the sense that 
$\nabla_{X_\tau} X_\tau =0$, where $\nabla_X$ is the covariant derivative with respect to 
$X$ induced by the metric $B$. 
This derivative is specified by the condition ${\rm L}_X B(X_1, X_2) = B(
\nabla_X X_1, X_2) + B(X_1, \nabla_X X_2)$, used to calculate the Christoffel symbols
$\Gamma$. Because ${\rm L}_{Z_i} \zeta_j(Z_k) = \epsilon_{ijk}$, in the basis 
$\{ Z_k, k=1,2,3 \}$ of the left-invariant fields we get 
$\nabla_{Z_j} Z_k = \sum_i \Gamma^i_{jk} Z_i$ with $\Gamma^i_{jk} = -(I_j -I_k) 
\epsilon_{ijk} /I_i$. Taking $X_\tau$ of the form  $X_\tau = \sum_k \omega'_k (t)  Z_k$, the 
condition $\nabla_{X_\tau} X_\tau =0$ yields $\dot{\omega}'_i = - \sum_{jk} \Gamma^i_{jk} 
\omega'_j \omega'_k$, which is the same as (\ref{eul}). For a spherical rotator 
$I_1=I_2=I_3 \equiv I $, $\Gamma =0$, 
\begin{equation}
T = \frac{I}{2} (\dot{\varphi}^2 +  \dot{\theta}^2  + \dot{\psi}^2 + 
2 \cos \theta ~\dot{\varphi} \dot{\psi} ) ~~, \label{trig1}
\end{equation}
and $\vec{\rho} = I \vec{\omega}'$ is a constant.  In the integrable basis $\partial_\varphi, 
\partial_\theta, \partial_\psi$, (\ref{trig1}) corresponds to a metric tensor $\hat{g}$ having the
non-vanishing components $\hat{g}_{\varphi \varphi} = \hat{g}_{\theta \theta} = \hat{g}_{\psi \psi} =1$, 
  $\hat{g}_{\varphi \psi} = \hat{g}_{\psi \varphi} = \cos \theta$, ($dv_R= d \varphi d \theta d \psi \sqrt{det \hat{g}}$), 
which yields the Ricci tensor $\hat{R} = \hat{g}/2$, and the scalar curvature 
\begin{equation}
R = Tr(\hat{g}^{-1} \hat{R} ) = \frac{3}{2}  ~~. \label{sc}
\end{equation}
\section{The classical coherent distributions}
Let $(M_\mu, \Omega_\mu)$  be the classical phase-space of an elementary rotator $\mu$ 
(e.g. molecule), and  $(M_\Gamma, \Omega_\Gamma) $ the phase-space of the ensemble (gas)  
consisting of $N$ identical elementary rotators \cite{somm}, 
$ M_\Gamma = M_1 \times M_2 \times ... M_N$, $\Omega_\Gamma = \sum_{\mu =1}^N  \Omega_\mu$.   
For a statistical description of the ensemble we define on each manifold $M_\mu$ 
a partition in $K$ infinitesimal cells, $\{ b_j~;~ j=1,K \}$,  of volume 
$\delta v^j = \int_{b_j} dv_R d^3 \rho$, 
\begin{equation}
 d v_R \equiv \sin \theta d \varphi d \theta  d \psi ~~,~~d^3 \rho \equiv d \rho_1
 d \rho_2 d \rho_3  ~~.    \label{evol}
 \end{equation} 
This partition induces a partition of  $M_\Gamma$ in  $n_B=K^N$ cells  $ B_j$  of volume  
$\delta  V_\Gamma^j$, $j=1,n_B$. Denoting by $w_j$ the probability of finding the representative
point $m \in M_\Gamma$, for the state of the ensemble  at the time $t$, localized in $B_j$, 
the ratio ${\sf F}_j = w_j  / \delta V_\Gamma^j$ defines the distribution function ${\sf F}$ 
of the probability density. This is symmetric at the permutation of the rotator indices and 
normalized by 
\begin{equation}
\int_{M_\Gamma} dV_\Gamma  ~{\sf F}  =1~~,~~  dV_\Gamma = \Pi_{\mu =1}^N (dv_R  d^3 \rho)_\mu ~~. 
\end{equation}
\indent 
It is important to remark that a definition of ${\sf F}$ with respect to the partition 
$\{ B_j~;~ j=1,n_B \}$ does not ensure ${\sf F} \in {\cal F}(M_\Gamma)$. A more suitable
description would require a covering of $M_\Gamma$, defined as an indexed system of 
open sets $\{ U_i, i \in I \}$, such that  $\cup_i U_i= M_\Gamma$, and a system of 
$1$-cochains \cite{fh}, associating to each pair of indices $i,j$ from $I$ a "transition" 
function $f_{(i,j)}$ on $U_i \cap U_j$, between ${\sf F}_i$ and ${\sf F}_j$. 
 \\ \indent 
Because $dv_R d^3 \rho \sim \vert \Omega^3 \vert $, the volume element
$ dV_\Gamma$ is invariant to the Hamiltonian flow on $M_\Gamma$, and ${\sf F}$
evolves according to the continuity (Liouville)
equation
\begin{equation}
\partial_t {\sf F}  = \{ H_\Gamma ,  {\sf F}  \}_\Gamma   ~~,  \label{lieq}
\end{equation}
where $\{~,~\}_\Gamma$ is the  Poisson bracket on $M_\Gamma$ and $H_\Gamma$ is the Hamiltonian of the 
ensemble. The mean value of an observable $A \in {\cal F}(M_\Gamma)$ is 
\begin{equation}
< A >_{\sf F} =   \int_{M_\Gamma} dV_\Gamma ~ {\sf F} ~A ~~, 
\end{equation}
and
\begin{equation} 
\frac{d <A>_{\sf F}}{dt} = < \{ A, H_\Gamma \}_\Gamma >_{\sf F}~~.   \label{da} 
\end{equation} 
For an ensemble of non-interacting rotators a particular solution of (\ref{lieq}) is
${\sf F}(m,t) = \Pi_{\mu=1}^N {\sf f}({\cal R}_\mu, \vec{\rho}_\mu,t)$, where ${\sf f} \in
{\cal F} (M_\mu)$ is a solution of the Liouville equation on $T^*SO(3, {\mathbb R})$,  
$\partial_t {\sf f}  = \{ H ,  {\sf f}  \}$,
with $H=\sum_k \rho_k^2/2I_k$, namely 
\begin{equation}
\partial_t {\sf f} + \sum_k \frac{\rho_k}{I_k} Z_k {\sf f} + \sum_{ijk} \epsilon_{ijk} 
\rho_i \frac{\rho_j}{I_j} \partial_{\rho_k} {\sf f} =0 ~~.  \label{leq0}
\end{equation}
Let $\tilde{\sf f}$ be the Fourier transform of ${\sf f}$ with respect to the intrinsic 
angular momentum, 
\begin{equation}
\tilde{\sf f}({\cal R}, {\bf r},t) =  \int d^3 \rho ~\re^{\ri {\bf r} \cdot \vec{\rho}}
~{\sf  f}({\cal R},\vec{\rho},t) ~~. \label{fr1}
\end{equation}
Thus, if  ${\sf  f}({\cal R},\vec{\rho},t)$ is a solution of (\ref{leq0}), 
then $\tilde{\sf  f}({\cal R},{\bf r},t)$ will satisfy  
\begin{equation}
\partial_t \tilde{\sf f} -  \ri \sum_k \frac{1}{I_k}  Z_k
\partial_{r_k} \tilde{\sf f} +
\ri \sum_{ijk}  \epsilon_{ijk} \frac{r_i}{I_k} \partial_{r_j} \partial_{r_k}
\tilde{\sf f} =0~~. \label{fle}
\end{equation}
Particular solutions of this equation are related to local Lagrangian submanifolds 
$\Lambda \subset T^*SO(3, {\mathbb R})$, such that $\Theta \vert_\Lambda = dS$, where 
$S({\cal R},t)$ is the generating function of the Hamilton-Jacobi theory. Presuming that
$\Lambda$ exists by the complete integrability of (\ref{eul}), these solutions  are of the form  
\begin{equation}
\tilde{\sf  f}_0({\cal R},{\bf r},t) = {\sf n}({\cal R},t) \re^{\ri {\bf r} 
\cdot {\bf Z} S}                                                            \label{f0r} 
\end{equation}
where the density\footnote{To ensure that ${\sf n}$ remains positive all the time, it should 
be taken of the form ${\sf n} = \Psi^* \Psi$, where $\Psi$ in general is complex.}
${\sf n} \ge 0$ and $S$ are real functions on $SO(3, {\mathbb R})$ which 
satisfy the continuity and, respectively, the Hamilton-Jacobi equations 
\begin{equation}
\partial_t {\sf n} + \sum_k Z_k ({\sf n} \frac{Z_k S}{I_k})  =0~~,~~
Z_i [ \partial_t S + \sum_k \frac{(Z_k S)^2}{2I_k}] =0 ~~.
\end{equation}
The inverse of (\ref{fr1}),  
\begin{equation}
{\sf f}({\cal R}, \vec{\rho},t) = \frac{1}{(2 \pi)^3} \int d^3 {\rm r} ~\re^{- \ri {\bf r} 
\cdot \vec{\rho}}~ \tilde{\sf  f}({\cal R},{\bf r},t)  \label{fr2}
\end{equation}
takes in this case the form of the classical "action distributions", 
\begin{equation}
 {\sf f}_0({\cal R}, \vec{\rho},t) = {\sf n}  \delta (\vec{\rho} - {\bf Z} S) ~~. \label{f0}
\end{equation}
These are coherent solutions of (\ref{leq0}) in the sense that during time evolution remain 
the same functionals of ${\sf n}$ and $S$. 
\section{Quantum phase-space distributions}
The quantum (Wigner-type) distributions are related to a peculiar form of 
$\tilde{\sf  f}({\cal R},{\bf r},t)$, in which the variable ${\bf r}$ (Fourier dual to 
$\vec{\rho}$) enters as a parameter for translations on the configuration space.   
To obtain this functional we note that  ${\bf Z}$ in (\ref{f0r}) 
is the generator of the translations to the right, and therefore  
\begin{equation}
 {\bf r} \cdot {\bf Z} S({\cal R})  = \lim_{\sigma \rightarrow 0 } \frac{1}{\sigma} 
[ S({\cal R} \re^{\sigma {\bf r} \cdot \vec{\xi}/2}) -
S({\cal R} \re^{- \sigma {\bf r} \cdot \vec{\xi}/2})]~~, \label{lim}
\end{equation}
where $\sigma$ is a real parameter, $\sigma {\bf r} \cdot \vec{\xi}$ is an
element of the Lie algebra ${\frak so}(3,{\mathbb R})$, and $\gamma = \sigma \vert 
{\bf r} \vert $ has the significance of a rotation angle. The "${\bf r}$-translations" will be 
defined using the first-order finite differences expression  for (\ref{lim}). \\ \indent 
A natural discretization of $SO(3, {\mathbb R})$ is provided by its discrete subgroups,
the largest covering uniformly $SO(3, {\mathbb R})$ being  the icosahedron group,
with 60 elements  \cite{su14}. Discretization  with statistical significance is the partition in
elementary cells (Section 3), while  for dynamics, we note that the commutation 
relations $[Z_i, Z_j ] = \epsilon_{ijk} Z_k$ are due to the nonlinearity of the operator 
${\bf Z}$, which is effective only outside a certain "infinitesimal domain" around identity. 
On the Cartan subalgebra of any semisimple Lie algebra we can introduce a lattice structure, 
dual to the lattice of the weights, while for the plane rotator discrete rotation angles are 
associated with a complete orthonormal set of angle states \cite{ao1,ao2}. Thus, near 
${\bf r} =0$ the derivative (\ref{lim}) may retain the nonlocal, finite differences expression, 
in which $\sigma$ is a small, but finite constant. Presuming that for $\sigma \rightarrow \hbar >0$ 
the 2-point derivative (\ref{lim}) should be  correlated with a 2-point form of the local entropy 
(or information) \cite{cdq2}, $\ln {\sf n}({\cal R}) \rightarrow [   \ln {\sf n}({\cal R}\re^{\hbar 
{\bf r} \cdot \vec{\xi}/2}) + \ln {\sf n}({\cal R}\re^{- \hbar {\bf r} \cdot \vec{\xi}/2}) ]/2$, 
the classical distribution (\ref{f0r}) takes  the "quantum"  form
\begin{equation}
\tilde{\sf  f}_q ({\cal R},{\bf r},t) = \Psi({\cal R} \re^{\hbar {\bf r} \cdot \vec{\xi}/2}) 
\Psi^* ({\cal R} \re^{- \hbar {\bf r} \cdot \vec{\xi}/2})~~,
\end{equation}
where $\Psi = \sqrt{\sf n}~ \re^{\ri S / \hbar}$ is the complex wave function. This expression,
in principle, will provide by (\ref{fr2}) a phase-space distribution  ${\sf f}_q$, but 
due to the infinite integration domain over ${\bf r}$, such a functional will not have the
properties of the Wigner quasiprobability distributions. Therefore, at $\rho / \hbar$ large,
we may consider instead the functional
\begin{equation}
{\sf f}_\Psi ({\cal R}, \vec{\rho},t) = \frac{1}{(2 \pi)^3} \int_{r \le \pi / \hbar} d^3 {\rm r} 
j_0^2(\frac{ \hbar r}{2}) ~\re^{- \ri {\bf r} \cdot \vec{\rho}} ~ 
\tilde{\sf  f}_q ({\cal R},{\bf r},t) ~~,  \label{wf}
\end{equation}
where $r = \vert {\bf r} \vert$ and $j_0(x) =x^{-1} \sin x$. Using the new variable
$\vec{\gamma} = \hbar {\bf r}$, (\ref{wf}) can also be written in the form
\begin{equation}
{\sf f}_\Psi ({\cal R}, \vec{\rho}) = \frac{1}{(2 \pi \hbar)^3} \int_{\gamma \le \pi} d^3 
\gamma  j_0^2(\frac{ \gamma }{2}) ~\re^{- \ri \vec{\gamma} \cdot \vec{\rho}/ \hbar}   
\Psi({\cal R} \re^{\vec{\gamma} \cdot \vec{\xi}/2}) 
\Psi^* ({\cal R} \re^{- \vec{\gamma} \cdot \vec{\xi}/2})  ~~.  \label{wf1} 
\end{equation}
The vector $\vec{\gamma} = (\gamma_1, \gamma_2, \gamma_3) =  \gamma {\bf g}$, $\vert {\bf g} \vert =1$,
yields the parametrization of the rotation matrices by the exponential map\footnote{
$\gamma$ provides the character $\chi_1({\cal R}) = Tr({\cal R}) = 
2 \cos \gamma +1$. In the basis $\partial_\alpha$, $\partial_\beta$, $\partial_\gamma$ the metric
tensor $\hat{g}$ is diagonal, with $\hat{g}_{\alpha \alpha} = \hat{g}_{\beta \beta} \sin^2 \beta$, 
$\hat{g}_{\beta \beta} = 4 \sin^2(\gamma /2) $, $\hat{g}_{\gamma \gamma} =1$.}, 
(Appendix 1), and  the factor $j_0^2(\gamma /2)$,  (found also in 
\cite{su14}-14.7),  was introduced such that $j_0^2(\gamma /2)  d \gamma_1 d \gamma_2 d 
\gamma_3$ becomes in the Euler parametrization the volume element $dv_R=\sin \theta d 
\varphi d \theta d \psi$. \\ \indent
Let  $\Psi \in L^2(SO(3, {\mathbb R}))$, $\langle \Psi \vert \Psi \rangle =1$, with 
$\langle \Psi_1 \vert \Psi_2 \rangle \equiv \int_{SO3} dv_R \Psi_1^*({\cal R}) \Psi_2 ({\cal R})$. 
In this case, ${\sf f}_\Psi \in {\cal F}(T^*SO(3, {\mathbb R})) $    defined by (\ref{wf}) has the  properties: \\
1.  ${\sf f}_\Psi$ is real and normalized, ${\sf f}_\Psi ={\sf f}_\Psi^*$,   
$\int_{T^*SO3} dv_R d^3 \rho~ {\sf f}_\Psi = \langle \Psi \vert \Psi \rangle =1 $. \\
2.  The phase-space overlap between ${\sf f}_{\Psi 1}$ and ${\sf f}_{\Psi 2}$ is    
\begin{equation}
< {\sf f}_{\Psi 1} >_{{\sf f}_{\Psi 2}} = \int_{T^*SO3} dv_R d^3 \rho~ {\sf f}_{\Psi 1} {\sf f}_{\Psi 2}= 
\vert \langle \Psi_1 \vert \Psi_2 \rangle \vert^2/ ( 2 \pi \hbar)^3 \ge 0 ~~,  
\end{equation}
such that ${\sf f}_\Psi \in L^2(T^*SO(3, {\mathbb R}))$. \\
3.  $ \lim_{\hbar \rightarrow 0} {\sf f}_\Psi =  {\sf f}_0 \ge 0$. The limit is the positive, 
coherent distribution for the classical Liouville equation, ${\sf f}_0= {\sf n}  
\delta (\vec{\rho} - {\bf Z} S)$.  \\ 
4. If $\Lambda_{\cal R} $  is the adjoint action of ${\cal R}= \re^{\vec{\gamma} \cdot \vec{\xi}} 
\in SO(3, {\mathbb R})$ by translations to the left (rotations of the laboratory frame), and
$\hat{U}_{\cal R}^\ell  =  \re^{\vec{\gamma} \cdot {\bf Y}}$, then 
$ \Lambda^*_{\cal R} {\sf f}_\Psi = {\sf f}_{\hat{U}_{\cal R}^\ell \Psi}$. \\  
5. The coordinate and momentum space distributions $F^{cs}_\Psi, F^{ms}_\Psi$ are  
\begin{equation}
F^{cs}_\Psi ({\cal R}) \equiv \int d^3 \rho {\sf f}_\Psi ({\cal R}, \vec{\rho}) = 
 \vert \Psi ({\cal R}) \vert^2~~,
\end{equation}
and  $F^{ms}_\Psi (\vec{\rho}) \equiv \int_{SO3} dv_R {\sf f}_\Psi ({\cal R}, \vec{\rho}) = 
\langle \Psi \vert \hat{P}_{\vec{\rho}} \vert \Psi \rangle$, where
\begin{equation}
\hat{P}_{\vec{\rho}} =\frac{1}{(2 \pi \hbar)^3}  \int_{\gamma \le \pi} d^3 \gamma j_0^2(\frac{ 
\gamma }{2})  \re^{\ri \vec{\gamma} \cdot (\hat{\bf L}' - \vec{\rho})/ \hbar}~~,  
\end{equation}
with $\hat{\bf L}' = - \ri \hbar {\bf Z}$ denoting the intrinsic angular momentum operator,
 Hermitian with respect to the volume element $dv_R$. Thus, $\hat{P}_{\vec{\rho}}$  is also 
Hermitian, commutes with  $(\hat{\bf L}')^2$, and if $\vec{\rho} = \rho {\bf e}_3'$, then 
$[ \hat{P}_{ \rho {\bf e}_3'} , \hat{L}'_3]=0$, namely $\hat{P}_{ \rho {\bf e}_3'}$ is 
diagonal in the angular momentum basis $\vert LK \rangle$.  For $\Psi_0= 1/\sqrt{8 \pi^2}$,  
$F^{ms}_{\Psi 0} (\vec{\rho})$ attains the maximum $8 \pi^2 /(2 \pi \hbar)^3$
at $\rho=0$, and then vanishes in an oscillatory manner as $\rho$ increases. 
 \\ \indent
These properties indicate that (\ref{wf}) provides a Wigner-type map from complex wave 
functions $\Psi \in L^2(SO(3, {\mathbb R}))$ to real quasiprobaility distributions
${\sf f}_\Psi$ on the classical phase-space $M=SO(3, {\mathbb R}) \times {\mathbb R}^3$.  
Concerning the observables of interest, the mean values of the intrinsic angular 
momentum components, 
\begin{equation}
< \rho_k >_{{\sf f}_\Psi}  =  \int_M  d v_R~ d^3\rho  ~\rho_k ~{\sf f}_\Psi = 
\end{equation}
$$
\frac{1}{(2 \pi \hbar)^3}  \int_M d v_R~d^3\rho~   \int_{\gamma \le \pi} d^3 
\gamma  j_0^2(\frac{ \gamma }{2}) ~ ( \ri \hbar \partial_{\gamma_k} 
\re^{- \ri \vec{\gamma} \cdot \vec{\rho}/ \hbar} )  
\Psi({\cal R} \re^{\vec{\gamma} \cdot \vec{\xi}/2}) 
\Psi^* ({\cal R} \re^{- \vec{\gamma} \cdot \vec{\xi}/2}) 
  $$  
can be easily estimated because the integral over $d^3 \rho$ yields a delta function
$\delta(\vec{\gamma})$, the first derivative of $j_0^2(\gamma/2)$ vanishes at $\gamma =0$,  
and $\partial_{\gamma_i} \Psi({\cal R} \re^{\vec{\gamma} \cdot \vec{\xi}/2}) 
\vert_{\gamma =0} =  Z_i \Psi({\cal R} ) /2$,  such that 
\begin{equation}
< \rho_k >_{{\sf f}_\Psi}  =  \frac{1}{2}  \int_{SO3} dv_R [ (\hat{L}_k' \Psi) \Psi^* + 
\Psi (\hat{L}_k' \Psi)^*] =  \langle \Psi \vert  \hat{L}_k' \vert \Psi \rangle ~~. 
\end{equation}
The calculation of  $< \rho_k^2 >_{{\sf f}_\Psi}$ involves second derivatives with respect
to $\gamma_k$ and proceeds similarly, excepting for the second derivative  
$\partial^2_{\gamma_k} j_0^2(\gamma/2 )$ at $\gamma=0$, which is $-1/6$. Thus, the operator
associated to $\rho^2_k$  by ${\sf f}_\Psi$  is $(\hat{L}_k')^2 + \hbar^2/6$, and the Hamiltonian
operator
\begin{equation}
\hat{H} = \sum_k( \frac{\hat{L'}_k^2}{2I_k} + \frac{ \hbar^2}{12 I_k} ) 
\end{equation}
contains beside the usual part a zero-point energy term $\epsilon_0=\sum_k \hbar^2 /12I_k$. 
This result is consistent with the expected dynamics, as a necessary condition 
for ${\sf f}_\Psi$ to be a solution of the Liouville equation is that  
$\ri \hbar \partial_t \Psi = \hat{H} \Psi$ (Appendix 3). Moreover,   for a spherical rotator 
$\epsilon_0 = \hbar^2 /4I = \hbar^2 R / 6I$, where $R$ is the scalar 
curvature (\ref{sc}), in agreement with \cite{gqa}.    
\\ \indent
The factor  $\re^{- \ri \vec{\gamma} \cdot \vec{\rho}/ \hbar}$ from (\ref{wf1}) 
can also be written  in terms of the 1-form 
$\Theta_\rho = \sum_k \rho_k \zeta_k$, where $\rho_k$ are constants. Because ${\bf g}
\cdot {\bf Z} = \partial_\gamma$, we get $ \vec{\gamma} \cdot \vec{\rho}  = 
\gamma \langle  \Theta_\rho, \partial_\gamma  \rangle = \hbar \Phi_\rho ({\cal R}_\gamma, 
{\cal L}_\gamma)$, where the phase
\begin{equation}
\Phi_\rho ({\cal R}_\gamma, {\cal L}_\gamma) = \frac{1}{\hbar} 
\int_{{\cal R}_\gamma^{-1/2}}^{{\cal R}_\gamma^{1/2}} \Theta_\rho \vert_{{\cal L}_\gamma}   
~~,~~ {\cal R}_\gamma = \re^{\vec{\gamma} \cdot \vec{\xi} } 
\end{equation}
is the integral of $\Theta_\rho/ \hbar$ along the line 
${\cal L}_\gamma =  \{ \re^{ t {\bf g} \cdot \vec{\xi} }~,~t \in [ - \gamma /2, 
\gamma /2] \}$ between ${\cal R}_\gamma^{-1/2}$ and ${\cal R}_\gamma^{1/2}$. An intrinsic
expression of ${\sf f}_\Psi$, independent of coordinates, should be therefore of the form
 \begin{equation}
{\sf f}_\Psi ({\cal R}, \vec{\rho}, [{\cal C}]) = \frac{1}{(2 \pi \hbar)^3} \int_{SO3} dv_{\tilde{R}} 
~\re^{- \ri \Phi_\rho (\tilde{\cal R}, {\cal C})  }   
\Psi({\cal R} \tilde{\cal R}^{1/2} ) 
\Psi^* ({\cal R} \tilde{\cal R}^{-1/2}  )  ~~,  
\end{equation}
where ${\cal C} \in [{\cal C}]$ is a curve of a suitable type from 
${\cal R}_\gamma^{-1/2}$ to ${\cal R}_\gamma^{1/2}$.  In this perspective, 
the results presented above correspond to the particular case in which 
 ${\cal C}$ is a  line, (geodesic for the metric $\hat{g}$), and may change as other 
options are considered. \\ \indent
One should note that $X_\rho = \vec{\rho} \cdot {\bf Z}$ is a characteristic 
vector of $\omega_\rho = - d \Theta_\rho$, ($ i_{X_\rho} \omega_\rho =0$), such that if
$G_\rho \subset SO(3, {\mathbb R})$ is the stability subgroup of $\vec{\rho}$, then
$M_\rho = SO(3, {\mathbb R}) / G_\rho$ has a symplectic structure given by the 2-form
$\omega_o$, $\pi^* \omega_o = \omega_\rho$, where $\pi$ is the projection 
$\pi : SO(3, {\mathbb R}) \rightarrow M_\rho$ along the orbits of $X_\rho$.   
The manifold  $M_\rho$ can be seen as the coadjoint orbit of $\vec{\rho}$, which is 
a sphere in ${\mathbb R}^3$ of radius $\rho$.  Wigner-type distributions  
on coadjoint orbits can be defined using the Plancherel transform \cite{afk}, 
but in the case of $SO(3, {\mathbb R})$  the result would be less general than ${\sf f}_\Psi 
\in {\cal F}(SO(3, {\mathbb R}) \times {\mathbb R}^3)$. 

\section{ Concluding remarks} 
The Wigner transform defines quasiprobability distributions on the phase-space 
$M=T^*{\mathbb R}^3$ which evolve as coherent solutions of the classical Liouville equation 
for  particles subject to uniform or elastic force fields \cite{cpw}. It also relates 
classical and quantum expectation values for many observables and provides a basis for 
the statistical interpretation of the scalar product between quantum wave functions. However,
to find the geometrical  structure underlying these properties, it is 
necessary to go beyond $T^*{\mathbb R}^3$. \\ \indent
The case $M=T^*SO(3, {\mathbb R})$ corresponds to the rigid rotator, and the proposed 
extension is ${\sf f}_\Psi$ of (\ref{wf}). This functional reduces in the limit 
$\hbar \rightarrow 0$  to the "action distribution" ${\sf f}_0$  (\ref{f0}), and ensures 
the usual quantization of the intrinsic angular momentum. Though, the expected value of 
the intrinsic Hamiltonian contains beside the usual "quantum part", a zero-point 
energy term. This term is positive definite and  reflects the compact, rather than the 
non-Abelian structure of $SO(3, {\mathbb R})$. \\ \indent
Formally, a local distribution function on  $T^*SO(3, {\mathbb R})$ can also be defined
as reduced Wigner transform on $T^*{\mathbb R}^4$, (Appendix 2). Compared to this, 
${\sf f}_\Psi$ has the advantage of providing more insight into the fundamental 
aspects of the transition between classical and quantum distributions, by taking into account 
the specific geometry of the problem. \\ 

\noindent
{\bf Appendix 1: $SO(3, {\mathbb R})$ parameters and  invariant vector fields} \\
Using the Euler parametrization\footnote{In quantum mechanics the second rotation is 
taken around the $Y$-axis, which corresponds to a change of $\varphi, \psi$ below into
$\varphi + \pi/2, \psi - \pi/2$.} a matrix ${\cal R} \in SO(3, {\mathbb R})$ can be 
expressed in the form 
 \begin{equation}
{\cal R} = \re^{ \varphi \xi_3} \re^{ \theta \xi_1} \re^{ \psi \xi_3}~~,  \label{gr} 
\end{equation}
where  $\xi_i~,i=1,2,3$, are 3 independent, antisymmetric,  $3 \times 3$ 
matrices, with elements  $(\xi_i)_{jk}  = - \epsilon_{ijk}$ 
and commutation relations   $[\xi_i, \xi_j] =  \epsilon_{ijk} \xi_k $.  
This definition corresponds to the adjoint action given by 
\begin{equation}
{\cal R} \xi_k  {\cal R}^{-1} =  \sum_i {\cal R}^{-1}_{ki} \xi_i ~~,   
\end{equation}
and to the action ${\bf e}_k' = {\cal R} {\bf e}_k= \sum_i {\cal R}^{-1}_{ki} {\bf e}_i$
on the laboratory frame versors ${\bf e}_k, k=1,2,3$, represented as column vectors ($3 \times 
1$ matrices),  such that ${\bf e}_k'$ are the versors of the rotated (intrinsic) frame. 
\\ \indent 
 A different parametrization is provided by the exponential map, 
$ {\cal R} = \re^{ \vec{ \gamma} \cdot \vec{ \xi }}$,  with 
\begin{equation}
\vec{\gamma} = (\gamma_1, \gamma_2, \gamma_3) =  \gamma {\bf g}~~,~~
{\bf g} = ( \cos \alpha \sin \beta, \sin \alpha \sin \beta, \cos \beta)~~, \label{gamma} 
\end{equation}
related to the Euler angles by 
\begin{equation}
\tan \varphi = - \frac{\cos \beta (1-\cos \gamma) \cos \alpha + \sin \gamma \sin \alpha}{
\cos \beta (1-\cos \gamma) \sin \alpha - \sin \gamma \cos \alpha}~~, \label{tphi}
\end{equation}
\begin{equation}
\tan \psi =  \frac{\cos \beta (1-\cos \gamma) \cos \alpha - \sin \gamma \sin \alpha}{
\cos \beta (1-\cos \gamma) \sin \alpha + \sin \gamma \cos \alpha}~~, \label{tpsi}
\end{equation}
$\cos \theta = 1 - 2 \sin^2 \beta \sin^2 (\gamma/2) $, and 
\begin{equation}
d \varphi =  d \alpha - a ( d \beta - \frac{ \cot \beta}{ \sin \gamma} d \gamma)~~,~~
d \psi =  - d \alpha - a ( d \beta - \frac{\cot \beta}{ \sin \gamma} d \gamma)~~,
\end{equation}
where $a = \sin \beta \sin \gamma /[2 - 2 \sin^2 \beta \sin^2(\gamma /2)]$.  \\ \indent
If   ${\cal R}$ depends on time, the derivative   
\begin{equation}
\dot{\cal R} = \sum_i \omega_i \xi_i {\cal R} = {\cal R} \sum_i \omega_i' \xi_i  \label{dr} 
\end{equation} 
defines the  components of the angular velocity $\omega_i$, $\omega_i'$, ($\omega_k'= 
\sum_i {\cal R}_{ki}^T  \omega_i$ ) in the laboratory, respectively in the  
intrinsic frame,
$$ \omega_1'= \dot{\theta} \cos \psi + \dot{ \varphi} \sin \theta \sin \psi $$
\begin{equation}
 \omega_2'= - \dot{\theta} \sin \psi + \dot{ \varphi} \sin \theta \cos \psi 
\end{equation}
$$ \omega_3'=   \dot{\psi}  + \dot{ \varphi} \cos \theta ~~. $$
Similarly to (\ref{dr}) it is convenient to define the generators of the translations to the 
left  ($Y_k$) and to the right ($Z_k$) by $Y_k {\cal R} =  {\xi}_k {\cal R}$, 
$ Z_k {\cal R}= {\cal R}  {\xi}_k $.  Explicitly, 
\begin{equation}
Y_1 = \cos \varphi \partial_\theta + \frac{\sin \varphi}{\sin \theta} ( \partial_\psi 
- \cos \theta \partial_\varphi ) ~,~
Y_2 = \sin \varphi \partial_\theta - \frac{\cos \varphi}{\sin \theta} ( \partial_\psi 
- \cos \theta \partial_\varphi ) ~,
\end{equation}
$Y_3 = \partial_\varphi$, and 
\begin{equation}
Z_1 = \cos \psi \partial_\theta + \frac{\sin \psi}{\sin \theta} ( \partial_\varphi 
- \cos \theta \partial_\psi )  ~,~
Z_2 =  - \sin \psi \partial_\theta + \frac{\cos \psi}{\sin \theta} ( \partial_\varphi 
- \cos \theta \partial_\psi ) ~,
\end{equation}
$Z_3 = \partial_\psi $. Thus, $Z_k(\varphi, \theta, \psi)  = - (-1)^k Y_k (\psi, \theta, \varphi) $ and
\begin{equation}
[Y_i, Y_j] = - \epsilon_{ijk} Y_k~~,~~ [Z_i, Z_j] =  \epsilon_{ijk} Z_k, ~~,~~ 
[Y_i, Z_j]=0~~,
\end{equation}
such that the components $x_k$ of a left-invariant vector field $X \in TSO(3, {\mathbb R})$, 
($[X, Y_k] =0$), $X= \sum_k x_k Z_k$, are interpreted as "intrinsic". \\ 

\noindent
{\bf Appendix 2: the $T^*{\mathbb R}^4 \rightarrow T^*SO(3, {\mathbb R})$ reduction} \\
The covering group of $SO(3, {\mathbb R})$ is $SU(2)$, which is homeomorphic to the unit sphere
$S^3 \subset {\mathbb R}^4$ \cite{ns}. Thus, a quasiprobability density on  $T^*SO(3, {\mathbb R})$ 
can be obtained by reduction from the standard Wigner transform of an extended wave function 
$\Psi_e(X)$ on ${\mathbb R}^4$,
\begin{equation}
{\sf f}_e (X,P) = \frac{1}{(2 \pi)^4} \int d^4 K ~\re^{- \ri K \cdot P} ~ 
\Psi_e (X +  \frac{\hbar}{2} K ) ~ \Psi_e^*(X- \frac{\hbar}{2} K) 
\end{equation}
where $X,P,K$ are 4-vectors of the form $X=(x_1,x_2,x_3,x_4)$, specified by the
Cartesian components $x_i$. Let $(r, \theta, \nu, \eta)$ be the spherical coordinates on
${\mathbb R}^4$, $\theta \in [0, \pi]$,  $~\nu, \eta \in [0, 2 \pi]$,  such that 
\begin{eqnarray}
x_1= r \cos \frac{ \theta}{2} \cos \nu ~~,~~ x_2 = r \cos \frac{ \theta}{2} \sin \nu  ~~ \label{x12} \\
x_3= r \sin \frac{ \theta}{2} \sin \eta ~~,~~ x_4 = r \sin \frac{ \theta}{2} \cos \eta ~~. \label{x34}  
\end{eqnarray}
Thus, $\sum_k x_k^2 =r^2$, and for $r=1$ we get  
\begin{equation}
\hat{a}_X = \left[ \begin{array}{cc}  x_1+ \ri x_2  & x_3 + \ri x_4 \\ 
- x_3 + \ri x_4   &  x_1 - \ri x_2  \end{array} \right] \in SU(2) ~~.  \label{ax} 
\end{equation}
The volume element 
$d^4 X \equiv d x_1 dx_2 dx_3 dx_4 =  r^3 \sin \theta d \theta d \nu d \eta d r/4$ 
yields for the ball of radius $R$ the 
volume $V_{B4} = \pi^2 R^4/2$. Thus, the "area" of the unit sphere $S^3$, which can be 
taken also as the volume of $SU(2)$, is $2 \pi^2$ (with this normalization the
volume of $SO(3, {\mathbb R})$ would be $\pi^2$).  \\ \indent
In spherical coordinates, the canonical symplectic form on $T^* {\mathbb R}^4$, 
$\omega_e = \sum_k dx_k \wedge dp_k $, $p_k = m \dot{x}_k$, has the form\footnote{For a 
particle of mass $m$ on $T^*{\mathbb R}^3$,  $\omega = \sum_k dx_k \wedge dp_k = dr \wedge 
dp_r + d \theta \wedge d p_\theta + d \varphi \wedge d p_\varphi$, with $p_r = m \dot{r}$, 
$p_\theta = m r^2 \dot{\theta}$, $p_\varphi = m r^2 \sin^2 \theta \dot{\varphi}$.}   
\begin{equation}
\omega_e = dr \wedge dp_r + d \theta \wedge dp_\theta + d \nu \wedge d p_\nu +   
d \eta \wedge d p_\eta~~, 
\end{equation}
\begin{equation}
p_r = m \dot{r}~~,~~p_\theta= m \frac{r^2}{4} \dot{\theta}~~,~~p_\nu = m r^2 \cos^2 
\frac{\theta}{2} \dot{\nu} ~~,~~p_\eta = m r^2 \sin^2 \frac{\theta}{2} \dot{\eta}~~,
\end{equation}
providing  the volume element   
$d^4Xd^4P = dr d \theta d \nu d \eta d p_r dp_\theta dp_\nu dp_\eta $. 
Because $\int d^4Xd^4P  ~{\sf f}_e (X,P) =1$, the reduced Wigner distribution
on $T^* SU(2)$ is defined by
\begin{equation}
{\sf f}_{SU2}(\theta, \nu, \eta, p_\theta, p_\nu, p_\eta)   = \int_0^\infty dr \int_{- \infty}^{\infty} dp_r ~~{\sf f}_e (X,P)~~.  \label{fsu2} 
\end{equation}
The projection $SU(2) \rightarrow SO(3, {\mathbb R})$ consists in the identification of
the $S^3$ poles $X=(1,0,0,0)$ and $(-1,0,0,0)$. Thus, it is reasonable to presume that ${\sf f}_{SU2}$ defines 
further a distribution ${\sf f}_{SO3}$ on $T^*SO(3, {\mathbb R})$, expressed in terms of the Euler angles $\theta, 
\varphi,  \psi$ and the related momenta $p_\theta, p_\varphi, p_\psi$, if  ${\sf f}_e$ is invariant at the inversion 
of the $x_1$ axis, $x_1 \rightarrow - x_1$. \\ \indent
The $\gamma$-coordinates (\ref{gamma}) can  be related to $X \in S^3$ by writing (\ref{ax}) 
as $\hat{a}_X = \re^{\ri \vec{\gamma} \cdot \vec{\sigma}/2} = \cos \gamma/2 + \ri {\bf g} \cdot
\vec{\sigma} \sin \gamma /2$, where $\gamma \in [0,2 \pi]$ and $\vec{\sigma}$ are the Pauli
matrices. The result $x_1 = \cos \gamma /2$, $x_2 = \cos \beta \sin \gamma/2$, 
$x_3 = \sin \alpha \sin \beta \sin \gamma /2$, $x_4 = \cos \alpha \sin \beta \sin \gamma /2$,
is consistent with (\ref{tphi}), (\ref{tpsi}) if $\theta$ in (\ref{x12}), (\ref{x34}) is the same as in 
(\ref{gr}), while $\nu = (\varphi + \psi)/2$, $\eta = (\varphi - \psi)/2$. Considering also 
$p_\nu = p_\varphi+ p_\psi$, $ p_\eta = p_\varphi - p_\psi$, (\ref{fsu2}) becomes a distribution
on  $T^*SO(3, {\mathbb R})$. The volume of the phase-space element is invariant at this change of coordinates, but  
the volume of the configuration space element decreases by a factor of 2. \\   

\noindent
{\bf Appendix 3: the coherence of the quantum distributions}  \\
{\it Theorem}. Let $\tilde{\sf f}_\Psi$ be the Fourier transform in angular momentum 
of the quantum quasiprobability distribution ${\sf f}_\Psi$ on $T^*SO(3, {\mathbb R})$,
\begin{equation}   
\tilde{\sf f}_\Psi({\cal R}, {\bf r},t) \equiv \int d^3 \rho ~\re^{\ri {\bf r } \cdot 
\vec{\rho} }~{\sf  f}_\Psi({\cal R},\vec{\rho},t) ~~. \label{a21}
\end{equation} 
Then, in the limit ${\bf r} \rightarrow 0$ : \\
{\it  i)}  $ \tilde{\sf f}_\Psi ({\cal R}, {\bf r}) \sim  
(\hat{U}_\gamma \Psi)_{ ({\cal R})} (\hat{U}_{- \gamma}  \Psi^*)_{ ({\cal R})} $,  $ 
\hat{U}_{\pm \gamma} =  \re^{\pm  \vec{\gamma} \cdot {\bf Z}/2} $, $\vec{\gamma} = \hbar {\bf r}$,    and \\
{\it ii)}   a necessary condition for ${\sf  f}_\Psi$ to satisfy the Liouville equation (\ref{leq0}) 
$\partial_t {\sf f}_\Psi  = \{ H ,  {\sf f}_\Psi  \}$ is that $\Psi$  evolves according to 
the time-dependent Schr\"odinger equation $\ri \hbar \partial_t \Psi = \hat{H} \Psi$.    
\\ \indent
{\bf Proof}. The first assertion was discussed in Section 4, at the definition of ${\sf f}_\Psi$. Thus, by the 
transform (\ref{a21}) of (\ref{wf}), excepting for the values $\vert {\bf r} \vert > \pi / \hbar$ when 
$\tilde{\sf f}_\Psi =0$,   we get
\begin{equation}
\tilde{\sf f}_\Psi ({\cal R}, {\bf r}) = j_0^2 (\gamma /2) 
\Psi({\cal R} \re^{\vec{\gamma} \cdot \vec{\xi}/2}) 
\Psi^* ({\cal R} \re^{- \vec{\gamma} \cdot \vec{\xi}/2})~~,~~
 \vec{\gamma} = \hbar {\bf r}~~,
\end{equation}
 in which $j_0^2(\gamma /2)$
decreases slowly from 1 to 0.4 when $\gamma \in [0, \pi]$.  \\ \indent 
For {\it ii)}, we start by using  (\ref{a21}) to obtain     
\begin{equation}
(\partial_t - \ri \hbar \hat{\bf a} \cdot \hat{\bf b})  \tilde{\sf f}_\Psi =  
\int d^3 \rho ~\re^{\ri \vec{\gamma} \cdot \vec{\rho} / \hbar}~ (\partial_t {\sf f}_\Psi  - 
\{ H ,  {\sf f}_\Psi  \})~~,  
\end{equation}
where $\hat{\bf a}$, $\hat{\bf b}$ are the operators 
$\hat{a}_i = I_i^{-1} \partial_{\gamma_i}$, $\hat{b}_i = Z_i - \sum_{jk} \epsilon_{ijk} 
\gamma_j \partial_{\gamma_k}$.
Thus, if ${\sf f}_\Psi$ is an exact (coherent) solution of (\ref{leq0}) then  
$(\partial_t - \ri \hbar \hat{\bf a} \cdot \hat{\bf b}) \tilde{\sf f}_\Psi =0$.
To calculate $\hat{b}_i \tilde{\sf f}_\Psi$ we note that ${\bf Z}$ acts only on the parameters 
of ${\cal R}$, such that   
\begin{equation}
\hat{\bf b} \tilde{\sf f}_\Psi = j_0^2(\gamma/2) [ (\hat{\bf b} \hat{U}_\gamma \Psi)
(\hat{U}_{-\gamma} \Psi^*) + (\hat{U}_{\gamma} \Psi) (\hat{\bf b} \hat{U}_{-\gamma} \Psi^*)] 
~~. \label{bi}
\end{equation}
Let ${\bf Y}^\gamma$ and ${\bf Z}^\gamma$ be the generators of the left and right translations 
for the element ${\cal R}_\gamma = \exp ( \vec{\gamma} \cdot \vec{\xi} )$, with 
$ \vec{\gamma} =  \gamma {\bf g}$.  Explicitly 
\begin{eqnarray}
Y^\gamma_i = \partial_{\gamma_i} + \frac{f-1}{\gamma} \nabla^A_i + \frac{1}{2}
({\bf g} \times \nabla^A )_i ~~, \\
Z^\gamma_i = \partial_{\gamma_i} + \frac{f-1}{\gamma} \nabla^A_i - \frac{1}{2}
({\bf g} \times \nabla^A )_i ~~,
\end{eqnarray}
where $f(\gamma) = (\gamma /2) \cot (\gamma /2)$ and $\nabla^A$ is the angular part
of $\partial_{\gamma_i} = {\bf g}_i \partial_\gamma + \nabla^A_i / \gamma $ (similar expressions 
have been derived in \cite{skm}). Thus, $\hat{\lambda} = {\bf g} \times \nabla^A = 
{\bf Y}^\gamma - {\bf Z}^\gamma$ is independent of $\gamma$ and acts only on ${\bf g}$  (the angular 
part of $\vec{\gamma}$,  configuration space for the simple rotator),
such that $\hat{\lambda}_i \hat{U}_{\pm \gamma} = Z_i \hat{U}_{\pm \gamma} -
 \hat{U}_{\pm \gamma} Z_i$. Because $\hat{b}_i = Z_i - \hat{\lambda}_i $, we get 
$ \hat{b}_i \hat{U}_{\pm \gamma} = (Z_i - \hat{\lambda}_i) \hat{U}_{\pm \gamma} = 
\hat{U}_{\pm \gamma} Z_i$, and (\ref{bi}) becomes 
\begin{equation}
\hat{b}_i \tilde{\sf f}_\Psi = j_0^2(\frac{\gamma}{2}) [ (\hat{U}_\gamma Z_i \Psi)
(\hat{U}_{-\gamma} \Psi^*) + (\hat{U}_{\gamma} \Psi) (\hat{U}_{-\gamma} Z_i \Psi^*) ]
~~.
\end{equation}
The action of $\hat{a}_i$ on this function takes a simple form only if $\vec{\gamma}=\hbar {\bf r}$ is 
small or along the $i$-axis, when $\hat{U}_{\gamma} = \re^{\gamma_i Z_i /2}$ and 
$\partial_{\gamma_i} \hat{U}_{\gamma} = (Z_i /2) \hat{U}_{\gamma} = \hat{U}_{\gamma} Z_i /2$.
Thus, presuming that $\gamma$ is small and  $\hat{a}_i \hat{U}_{\pm \gamma} \sim  \pm \hat{U}_{\pm 
\gamma} Z_i / 2I_i$,  we get
\begin{equation}
- \hbar^2 \hat{\bf a} \cdot \hat{\bf b} \tilde{\sf f}_\Psi \sim  
[ (\hat{U}_\gamma \hat{H} \Psi) (\hat{U}_{-\gamma} \Psi^*) - (\hat{U}_{\gamma} \Psi) 
(\hat{U}_{-\gamma} \hat{H} \Psi^*) ]~~,
\end{equation}
with $\hat{H} = - \hbar^2 \sum_i Z_i^2 / 2I_i$ (up to  a constant). Introducing 
$\hat{\Lambda} = \partial_t + \ri \hat{H} / \hbar$,      
\begin{equation}
(\partial_t - \ri \hbar \hat{\bf a} \cdot \hat{\bf b}) \tilde{\sf f}_\Psi \sim  
[ (\hat{U}_\gamma \hat{\Lambda} \Psi) (\hat{U}_{-\gamma} \Psi^*) + (\hat{U}_{\gamma} \Psi) 
(\hat{U}_{-\gamma} \hat{\Lambda}^* \Psi^*) ]~~,
\end{equation}
such that if ${\sf f}_\Psi$ is a solution of the classical Liouville equation then 
$\hat{\Lambda} \Psi =0$, namely $\Psi$ is a solution of the time-dependent 
Schr\"odinger equation $\ri \hbar \partial_t \Psi = \hat{H} \Psi$. In general,
we may expect that the coherence (functional form) of ${\sf f}_\Psi$ is maintained as long as 
$\tilde{\sf f}_\Psi ({\cal R}, {\bf r})$ has significant, non-vanishing values only for 
$\hbar \vert {\bf r} \vert \ll \pi $.  

\end{document}